\documentclass[final,3p,times,twocolumn]{elsarticle}

\usepackage{amssymb}
\usepackage{subfigure}
\usepackage{float}
\usepackage{color}

\usepackage{lineno, hyperref}

\journal{Nuclear Inst. and Methods in Physics Research, A}

\begin{document}

\begin{frontmatter}



\title{Low-Energy X-ray Performance of SOI Pixel Sensors for Astronomy, ``XRPIX''}


\author[kyoto]{Ryota Kodama}
\ead{kodama.ryota.62u@st.kyoto-u.ac.jp}
\author[kyoto]{Takeshi Go Tsuru} \author[kyoto]{Takaaki Tanaka} \author[kyoto]{Hiroyuki Uchida} \author[kyoto]{Kazuho Kayama} 
\author[kyoto]{Yuki Amano}
\author[miyazaki]{Ayaki Takeda} \author[miyazaki]{Koji Mori} \author[miyazaki]{Yusuke Nishioka} \author[miyazaki]{Masataka Yukumoto} \author[miyazaki]{Takahiro Hida}
\author[kek]{Yasuo Arai} \author[kek]{Ikuo Kurachi}
\author[tokyorika]{Takayoshi Kohmura} \author[tokyorika]{Kouichi Hagino} \author[tokyorika]{Mitsuki Hayashida} \author[tokyorika]{Masatoshi Kitajima}
\author[shizuoka]{Shoji Kawahito} \author[shizuoka]{Keita Yasutomi}
\author[okinawa]{Hiroki Kamehama}

\address[kyoto]{Department of Physics, Graduate School of Science, Kyoto University, Kitashirakawa Oiwake-cho, Sakyo-ku, Kyoto 606-8502, Japan}
\address[miyazaki]{Department of Applied Physics, Faculty of Engineering, University of Miyazaki,1-1 Gakuen Kibana-dai Nishi, Miyazaki 889-2192, Japan}
\address[kek]{Institute of Particle and Nuclear Studies, High Energy Accelerator Research Org., KEK, 1-1 Oho, Tsukuba 305-0801, Japan}
\address[tokyorika]{Department of Physics, Faculty of Science and Technology, Tokyo University of Science, 2641 Yamazaki, Noda, Chiba 278-8510, Japan}
\address[shizuoka]{Research Institute of Electronics, Shizuoka University, Johoku 3-5-1, Naka-ku, Hamamatsuu, Shizuoka 432-8011, Japan}
\address[okinawa]{National Institute of Technology, Okinawa College, Henoko 905, Nago-shi, Okinawa 905-2192, Japan}

\begin{abstract}
We have been developing a new type of X-ray pixel sensors, ``XRPIX", allowing us to perform imaging spectroscopy in the wide energy band of 1--20$ ~\mathrm{keV}$ for the future Japanese X-ray satellite ``FORCE".
The XRPIX devices are fabricated with complementary metal-oxide-semiconductor
silicon-on-insulator technology, and have the ``Event-Driven readout mode", in which only a hit event is read out by using hit information from a trigger output function equipped with each pixel. 
This paper reports on the low-energy X-ray performance of the ``XRPIX6E'' device with a Pinned Depleted Diode (PDD) structure. 
The PDD structure especially reduces the readout noise, and hence is expected to largely improve the quantum efficiencies for low-energy X-rays.
While F-K X-rays at $0.68~\mathrm{keV}$ and Al-K X-rays at $1.5~\mathrm{keV}$ are successfully detected in the ``Frame readout mode", in which all pixels are read out serially without using the trigger output function, the device is able to detect Al-K X-rays, but not F-K X-rays in the Event-Driven readout mode. 
Non-uniformity is observed in the counts maps of Al-K X-rays in the Event-Driven readout mode, which is due to region-to-region variation of the pedestal voltages at the input to the comparator circuit. 
The lowest available threshold energy is $1.1~ \mathrm{keV}$ for a small region in the device where the non-uniformity is minimized. 
The noise of the charge sensitive amplifier at the sense node and the noise related to the trigger output function are $\sim 18~e^{-}$ (rms) and $\sim 13~e^{-}$ (rms), respectively.

\end{abstract}




\end{frontmatter}


\section{Introduction}
We have been developing the future Japanese X-ray satellite ``FORCE",
whose main mission is to hunt for missing black holes in various mass scales 
from stellar-mass black holes to supermassive black holes 
and to trace their cosmic evolution\cite{Mori et al.(2016), Nakazawa et al.(2018) FORCE}. 
FORCE provides broadband imaging spectroscopy in 1--79~keV with an angular resolution better than $15''$ in a half-power diameter. 
For this purpose, the FORCE satellite has two identical pairs of a super mirror optics and a detector. 
As the optics, we employ light-weight Si mirrors provided by NASA's GSFC\cite{Zhang et al.(2019)}.
We adopt a hybrid type of detector consisting of Si and CdTe layers to cover the broadband, 
whose overall design is based on the hard X-ray imager (HXI) onboard the Hitomi satellite\cite{Nakazawa et al.(2018) HXI}. 
The Si sensor is required to cover the energy band from 1 to 20~keV, 
while a CdTe double-sided strip detector covers 20--79~keV. 

As the Si sensor for the FORCE mission, we have been developing a new type of X-ray sensors, called ``XRPIX'' in order to realize the wide-band observation\cite{Tsuru et al.(2018)}.
The XRPIXs are active pixel sensors based on silicon-on-insulator (SOI) 
and complementary metal-oxide-semiconductor (CMOS) technologies. 
The SOI pixel sensors are fabricated by sandwiching a buried oxide (BOX) layer between a low-resistivity Si circuit layer and a high-resistivity Si sensor layer\cite{Arai et al.(2011)}.
This unique structure allows us both high-speed readout CMOS electronics and a thick depletion layer to detect X-rays with a high sensitivity.

\begin{figure}[tbp]
	\centering
	\includegraphics[width=.95\columnwidth]{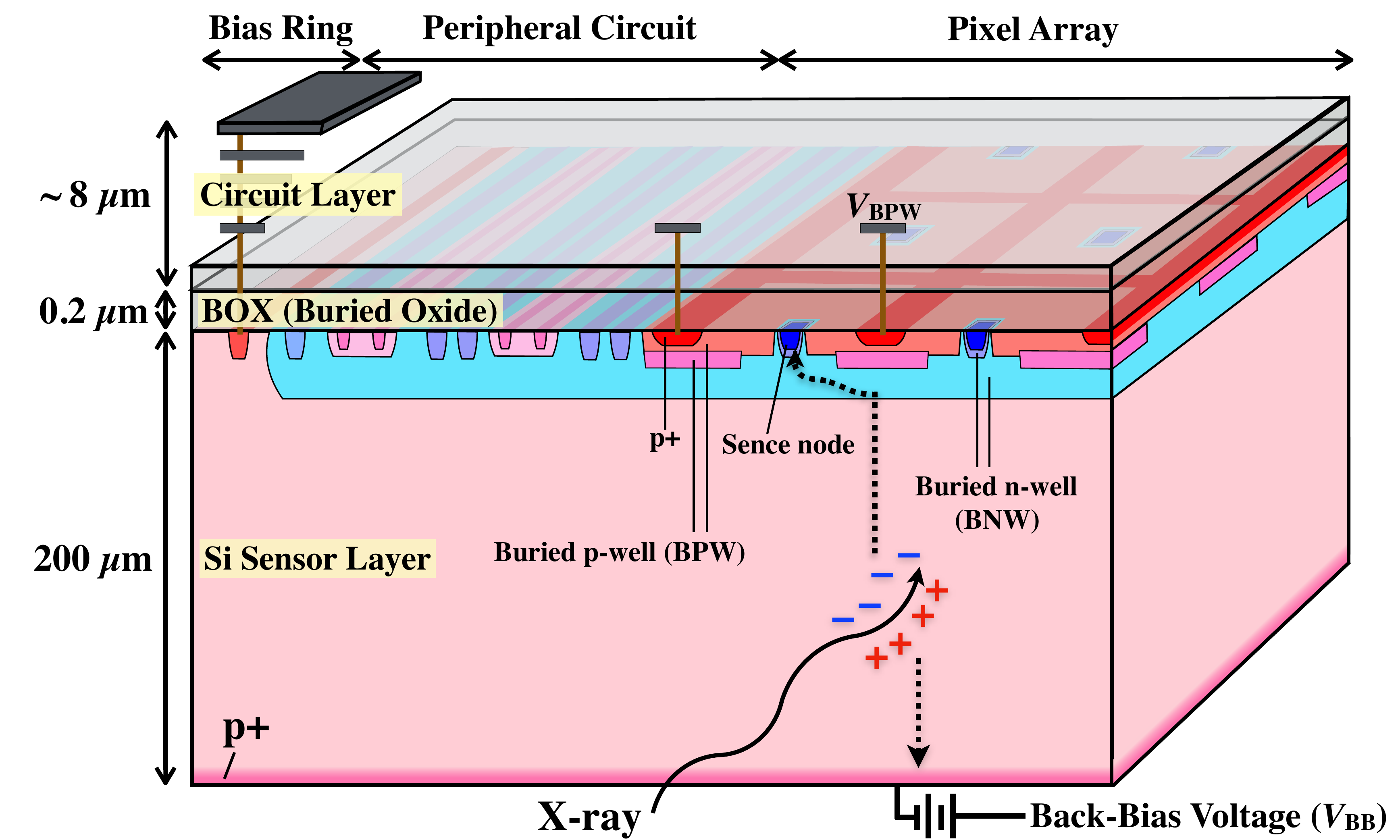}
	\caption{Cross-sectional view of XRPIX6E.}
	\label{fig:XRPIX6E}
\end{figure}

Each pixel in an XRPIX device is equipped with an event trigger output function with a time resolution better than 10~$\mu$sec. 
This time resolution enables us to observe fast time variation of X-ray emitting objects and to apply an anti-coincidence technique to reduce the non-X-ray background, which limits the sensitivity especially in the energy band above $\sim 5~\mathrm{keV}$. 
It also allows us to read out only the hit pixels, which mitigates pile-up for bright objects. 
We refer to this readout using the trigger output function as ``Event-Driven readout mode", which is the main mode in the FORCE mission. 
In addition, XRPIX devices have ``Frame readout mode'' as an optional mode, in which all pixels are read out serially after each timed exposure without using the trigger output function. 

Currently, one of our major development items of XRPIX is the performance for low-energy X-rays. 
Backside illumination (BI) is adopted in order to realize higher sensitivity for low-energy X-rays.
The scientific objectives of the FORCE mission require the XRPIX device to have an energy threshold lower than $1~\mathrm{keV}$ and an effective thickness of the dead layer of $ < 1~\mathrm{\mu m}$ including an Al optical blocking layer (OBL). 
Here the ``effective'' thickness is the one measured with X-ray detection and is often different from the one defined in terms of semiconductor processes. 
So far, we have achieved a performance of detection of $2.1~\mathrm{keV}$ X-rays and an effective thickness of 0.9--1~$\mathrm{\mu m}$ in the Frame readout mode\cite{Itou et al.(2016), Negishi et al.(2019)}. 
However, the required threshold of $1~\mathrm{keV}$ has not been reached yet. 
Furthermore, the evaluation was not done in the Event-Driven readout mode. 

Itou~et~al. (2016) pointed out that it is necessary to lower the readout noise to improve the low-energy X-ray performance and to reduce the effective thickness of the dead layer\cite{Itou et al.(2016)}. 
After the evaluations of XRPIX2b and XRPIX3b\cite{Itou et al.(2016), Negishi et al.(2019)}, we developed the new device, XRPIX6E, which successfully reduces the readout noise to $\sim 20~e^{-}$ in the Event-Driven readout mode by introducing the new device structure, Pinned Depleted Diode (PDD)\cite{Kamehama et al.(2018), Harada et al.(2019), Kayama et al.(2019)}. 
Thus, we also expect better performance for low-energy X-rays than XRPIX2b and XRPIX3b.

We here report on the performance of XRPIX6E for low-energy X-rays.
A device description of XRPIX6E is given in section \ref{sec:device}.
A low energy X-ray performance in the Frame readout mode for XRPIX6E is presented and discussed in section \ref{sec:QE}.
Then, section \ref{sec:trigger} is devoted for the performance in the Event-Driven readout mode.
Finally, conclusions are given in section \ref{sec:conclusion}.


\section{Device Description}
\label{sec:device}

\begin{figure}[tbp]
	\centering
	\includegraphics[width=.95\columnwidth]{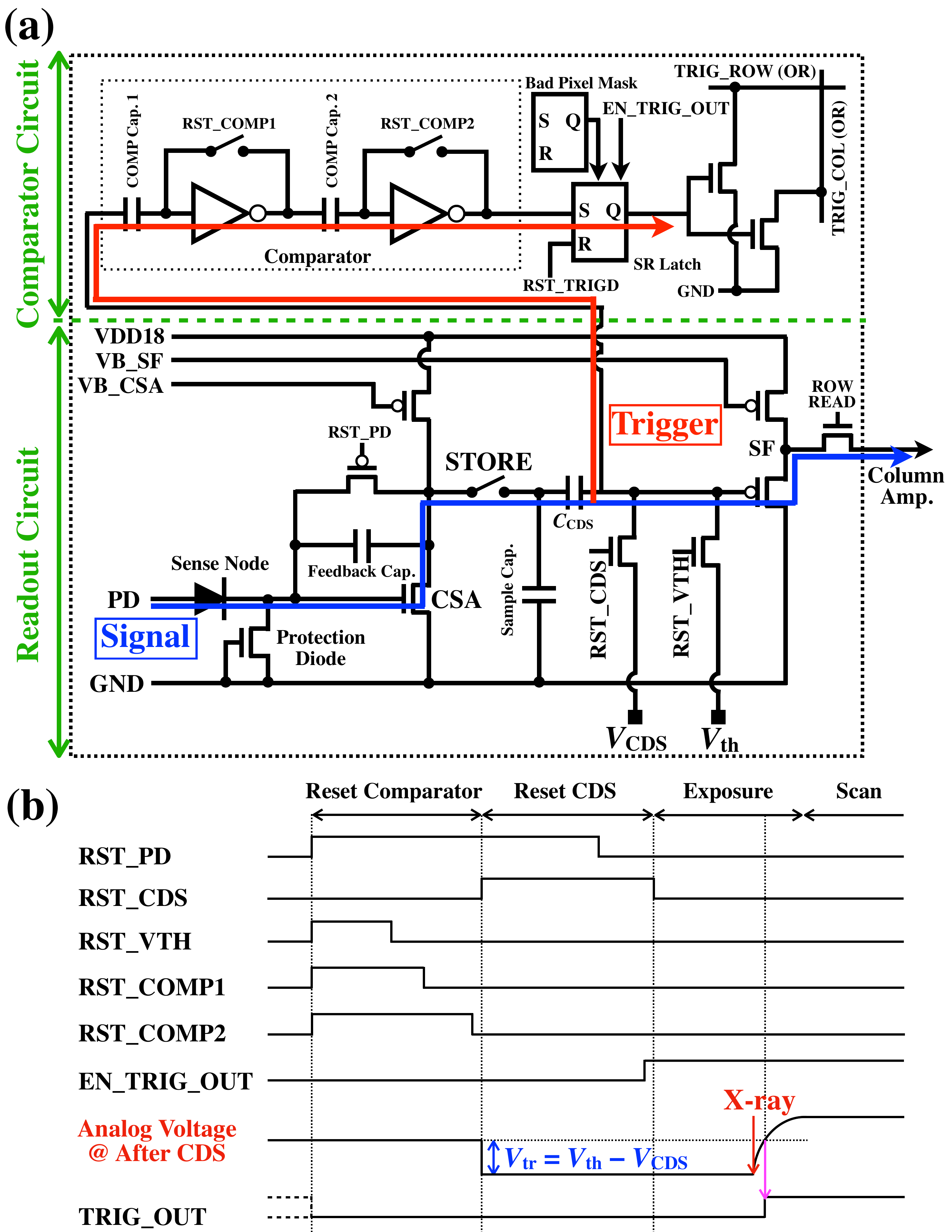}
	\caption{(a) Schematic of the pixel circuit of XRPIX6E. 
		It consists of ``readout circuit" and ``comparator circuit". 
		(b) Timing diagram of the readout sequence in the Event-Driven readout mode.}
	\label{fig:Circuit_and_Diagram}
\end{figure}

\begin{figure*}[htb]
	\centering{%
	\includegraphics[width=1.6\columnwidth]{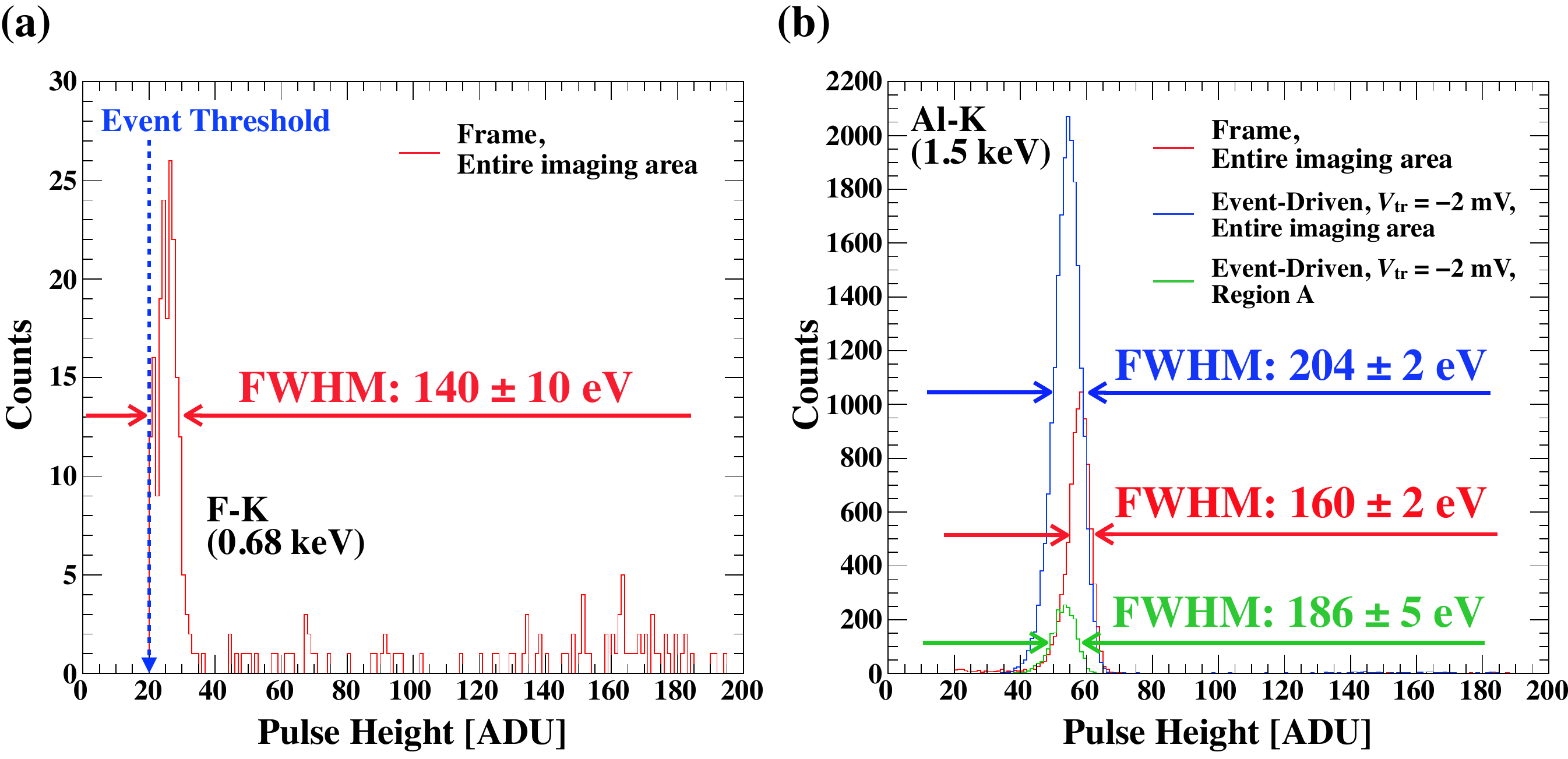}}
	\caption{X-ray spectra of single-pixel events obtained with XRPIX6E. 
	(a) A spectrum of F-K line in the Frame readout mode. 
	(b) Spectra of Al-K line obtained from the entire imaging area in the Frame readout mode (red), 
	and from the entire imaging area (blue) and  Region A (green) in the Event-Driven readout mode.}
	\label{fig:spectra_Al_F}
\end{figure*}

Fig.~\ref{fig:XRPIX6E} shows a cross-sectional view of the XRPIX6E.
The XRPIX6E device has $48~\times ~ 48$ pixels and a pixel size of $36~\mathrm{\mu m}~\times ~ 36~\mathrm{\mu m}$\cite{Harada et al.(2019), Kayama et al.(2019)}.
The sensor layer has a thickness of $200~\mathrm{\mu m}$ using a p-type floating zone wafer with a resistivity of $>25~\mathrm{k\Omega \cdot cm}$. 
A p$+$ layer with a thickness of $0.2~\mathrm{\mu m}$ as an X-ray entrance window and thin electrode to apply a back-bias voltage is formed on the backside of the device.
This process was followed by laser annealing. 
No Al optical blocking layer is deposited on the backside. 
The PDD structure is formed in the sensor layer at the interface with the BOX layer in order to reduce parasitic capacitance at the sense node and to suppress the electrical interference between the sense nodes and circuit layer by pinning the potential of the highly-doped buried p-well ($V_{\mathrm{BPW}}$ in Fig.~\ref{fig:XRPIX6E}) \cite{Kamehama et al.(2018)}. The PDD structure also improves the radiation hardness of the  transistors in the circuit layer. Applying an appropriate voltage to the $V_{\mathrm{BPW}}$ can eliminate the  back-gate effect on the transistors caused by charge build-up in the BOX layer due to the ionizing radiation.

Fig.~\ref{fig:Circuit_and_Diagram} shows the pixel circuit and timing diagram of XRPIX6E.
The basic configuration of the pixel circuit of the XRPIX6E device is the same as that of our previous devices except for a few changes\cite{Takeda et al.(2013)}. 
The pixel circuit consists of a readout circuit and a comparator circuit as shown in Fig.~\ref{fig:Circuit_and_Diagram}~(a). 
Signal charge generated by an X-ray is converted to a signal voltage by the charge sensitive amplifier (CSA) circuit. 
In the Frame readout mode, the signal voltage is read out through the pixel correlated double sampling (CDS) circuit and the pixel output source follower without using the comparator circuit. 
In the Event-Driven readout mode, we set the threshold voltage, V\_CDS, into the comparator circuit using RST\_CDS, and then start an exposure. 
As the comparator circuit detects that the signal voltage due to X-ray detection has exceeded the threshold voltage, the comparator circuit outputs a digital trigger signal to an external circuit as an event timing signal. 
After that, we read out the analog signal voltage of the triggering pixel and the $8\times 8$ pixels having the triggering pixel at their center as an X-ray event.

\section{QEs for Low-Energy X-rays and Dead-Layer Thickness in Frame Read Out Mode}
\label{sec:QE}

In this section, we evaluate the effective thickness of the dead layer by measuring the quantum efficiencies (QEs) for X-rays. 
Using the experimental setup given in Itou~et~al. (2016)\cite{Itou et al.(2016)}, we illuminated the XRPIX6E device with the three fluorescent X-ray emission lines of F-K ($0.68 ~ \mathrm{keV}$), Al-K ($1.5 ~ \mathrm{keV}$), Ti-K ($4.6 ~ \mathrm{keV}$). 
We also used a $^{57}\mathrm{Co}$ radioisotope to illuminate the device with Fe-K ($6.4 ~ \mathrm{keV}$ and $7.1 ~ \mathrm{keV}$) and the deexcitation line at $14.4 ~ \mathrm{keV}$ from $^{57}\mathrm{Fe}$ nuclei. 
The XRPIX6E device was cooled to $\mathrm{-60^\circ C}$ to reduce the dark current, and a back-bias voltage of $V_{\mathrm{BB}} = -200~\mathrm{V}$ was applied to fully deplete the sensor layer.
The data were read out in the Frame readout mode and analyzed in the same method described in Ryu~et~al. (2011) and Itou~et~al. (2016)\cite{Ryu et al.(2011), Itou et al.(2016)}. 
We collected X-ray events by extracting pixels whose pulse heights are higher than the event threshold and each pulse height of all 8 adjacent pixels, 
and then classified each X-ray event into one of the following grades of single-pixel, double-pixel, and others according to the pattern of whether or not each pulse height of 8 pixels next to the event center pixel exceeds the split threshold.
The readout noise of the CSA circuit operating in the Frame readout mode was $\sigma = 12~e^{-} ~(45 ~ \mathrm{eV})$, and
the event threshold and the split threshold were set to  $10~\sigma ~ (120 ~ e^{-})$ and $3~\sigma ~ (36 ~ e^{-})$, respectively.
Fig.~\ref{fig:spectra_Al_F} shows F-K and Al-K spectra of single-pixel events. 
This is the first clear detection of the F-K X-rays in the XRPIX series. 
The energy resolutions for the F-K and Al-K X-rays are $140 \pm 10~\mathrm{eV}$ and $160 \pm 2~\mathrm{eV}$ in Full Width at Half Maximum (FWHM), respectively.

\begin{figure}[tbp]
\centering
\includegraphics[width=.95\columnwidth]{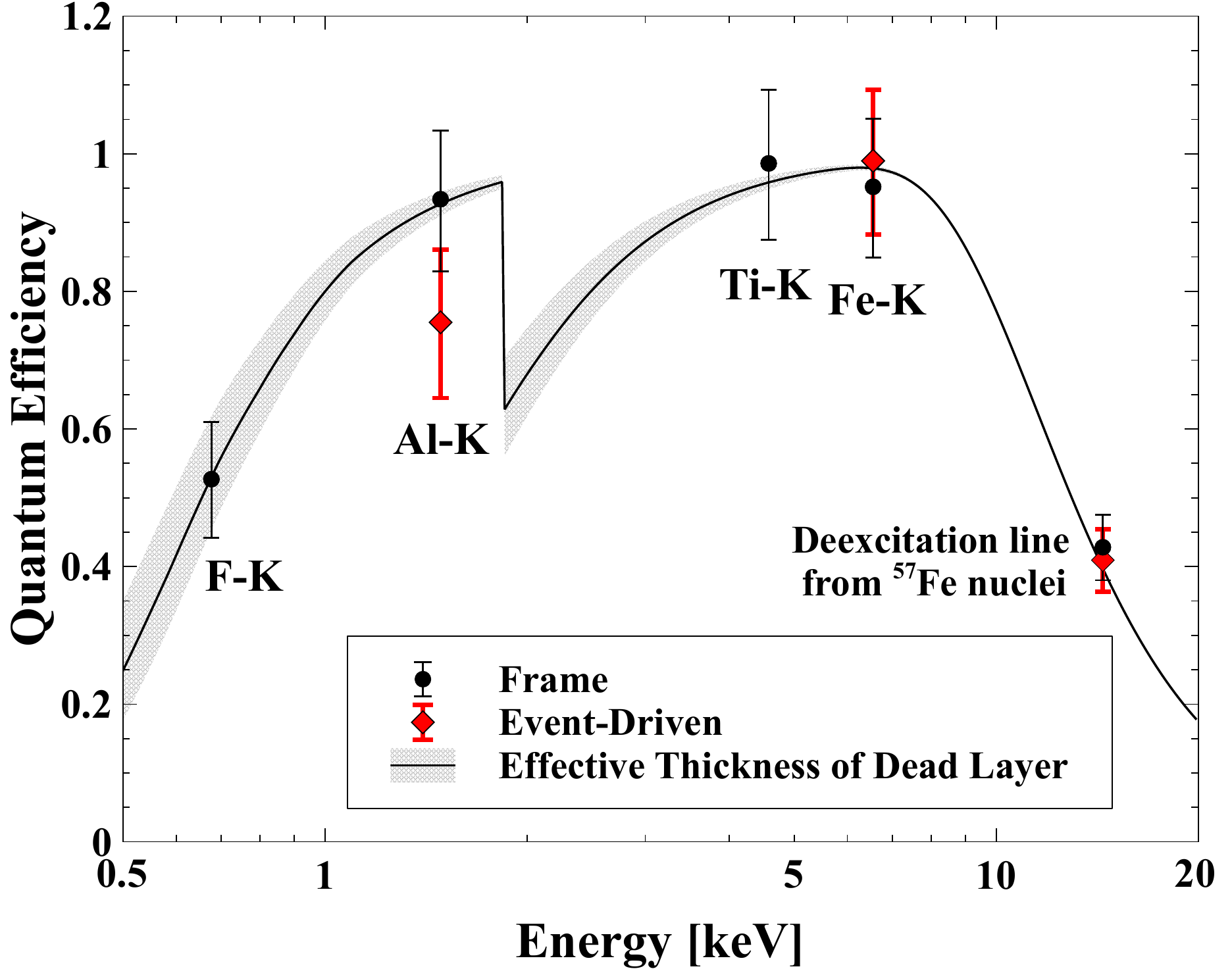}
\caption{Quantum efficiency (QE) as a function of the X-ray energy evaluated in the Frame and Event-Driven readout modes. 
	The gray shaded area corresponds to $\pm 1 \sigma$ error of the effective thickness of the dead layer calculated from QEs in the Frame readout mode.}
\label{fig:QE}
\end{figure}

Fig.~\ref{fig:QE} presents the QEs for all X-ray events, i.e. the sum of the events of the three grades.
The QE at $14.4~\mathrm{keV}$ agrees well with the expectation from the full depletion of the 200-$\mathrm{\mu m}$ thick sensor layer.
The QEs were fitted with the following function of X-ray energy
\[
\label{eq:QE}
\mathrm{QE} = \exp \left( -\frac{d}{\lambda_{\mathrm{Si}}(E)} \right) \left\{ 1 - \exp \left( -\frac{W - d}{\lambda_{\mathrm{Si}}(E)}\right) \right\},
\]
where $d$ is the effective thickness of the dead layer, $\lambda_{\mathrm{Si}}(E)$ is the attenuation length of Si at the X-ray energy of $E$, and $W = 200~\mathrm{\mu m}$ is the thickness of the sensor layer.
The effective thickness of the dead layer is estimated to be $d = 0.60 \pm 0.15 ~\mathrm{\mu m}$.
The error is dominated by the systematic uncertainty in the collimator size of the silicon drift detector, Amptek XR-100,  used for the calibration of the absolute flux of each fluorescent X-ray line\cite{Itou et al.(2016)}.
The FORCE requirement is $\mathrm{QE}~(= 0.688)$ at $1~\mathrm{keV}$ corresponding to the dead layer of Si with the  effective thickness of $1~\mathrm{\mu m}$. 
If the Al OBL thickness of $0.2~\mathrm{\mu m}$ having the transmittance of 0.938 at $1~\mathrm{keV}$, which is not processed on this device, is included in the dead layer, the transmittance of the remaining Si dead layer is required to be $0.688 / 0.938 = 0.733$.
It is translated to a Si thickness of $0.83~\mathrm{\mu m}$.
The measured value $d = 0.60 \pm 0.15 ~\mathrm{\mu m}$ satisfies this requirement. 

The electrode layer on the backside has a physical thickness of $\mathrm{0.2 ~\mu m}$.
This is rather thicker than that of the XRPIX2b-CZ-PZ ($\mathrm{0.1 ~\mu m}$) investigated by Itou et al. (2016)\cite{Itou et al.(2016)}. 
Nevertheless, the effective thickness of the dead layer of XRPIX6E obtained above is thinner than that of the XRPIX2b-CZ-PZ device. 
As discussed in Itou~et~al. (2016)\cite{Itou et al.(2016)}, this is because XRPIX6E has a lower readout noise than XRPIX2b-CZ-PZ ($61~e^{-}$ in rms).
This suggests that further improvement is possible with an even lower readout noise.
Assuming that signal charge from a F-K X-ray at $0.68 ~ \mathrm{keV}$ is shared evenly by two pixels, the signal charge collected in one pixel is $93 ~ e^{-}$. 
Such an event results in being undetected since the signal charge is below the event threshold of 
$120 ~ e^{-}$ (defined as $10~\sigma$ of the readout noise) that we applied here.
We need to reduce the readout noise even for efficiently detecting X-rays at $1~\mathrm{keV}$.
Let us consider a case where signal charge of an 1-keV X-ray is evenly shared by four pixels.
Signal charge per pixel is $68 ~ e^{-}$, which should be higher than the event threshold, or $10~\sigma$ of the readout noise.
Thus, the readout noise is required to be lower than $68/10 ~ e^{-} \sim 7 ~ e^{-}$. 

\section{Low-Energy X-ray Trigger Performance in Event-Driven Readout Mode}
\label{sec:trigger}

\begin{figure*}[htb]
	\centering{%
	\includegraphics[width=140mm]{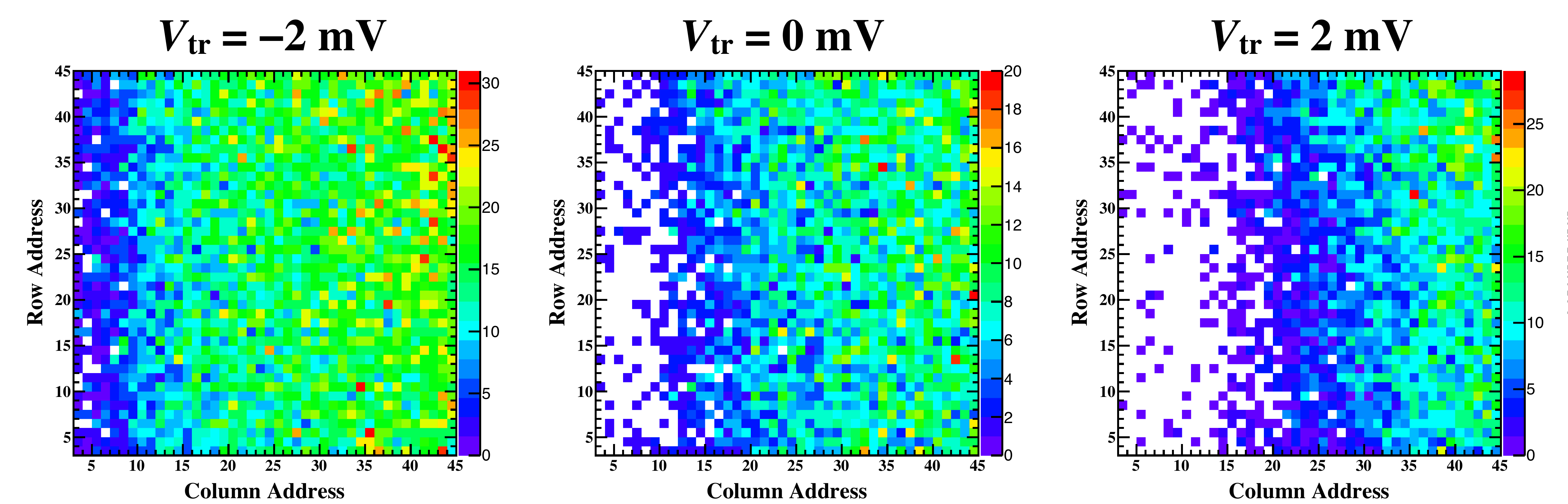}}
	\caption{Counts maps of trigger outputs for Al-K X-ray line. 
	Those obtained with $V_{\mathrm{tr}} = -2, 0,$ and $2 ~ \mathrm{mV}$ are shown from left to right.}
	\label{fig:Trigger_Map}
\end{figure*}

The physical thickness of the dead layer and the readout noise are the main factors to limit the QEs for low-energy X-rays in the Frame readout mode as discussed in the previous section. 
In the Event-Driven readout mode, on the other hand, the trigger performance also limits the QEs.
In this section, we investigate the trigger performance for low-energy X-rays in the Event-Driven readout mode.

\subsection{Non-uniformity in count map for low-energy X-rays}
\label{subsec:Non-uniformity}
We obtained counts maps of Al-K X-rays with various trigger threshold voltage set into the comparator circuit as shown in Fig.~\ref{fig:Trigger_Map}. 
Even though the entire imaging area was uniformly irradiated with X-rays, the counts maps show gradation.
As the threshold voltage is increased, the number of detected events, i.e. triggered events, decreased in the left region (lower column address) of the maps. 
Such non-uniformity of QEs or X-ray pulse heights is not observed in the Frame readout mode
since the pedestal levels of each pixel are subtracted in the data analysis.

\begin{figure}[htbp]
\centering
\includegraphics[width=.95\columnwidth]{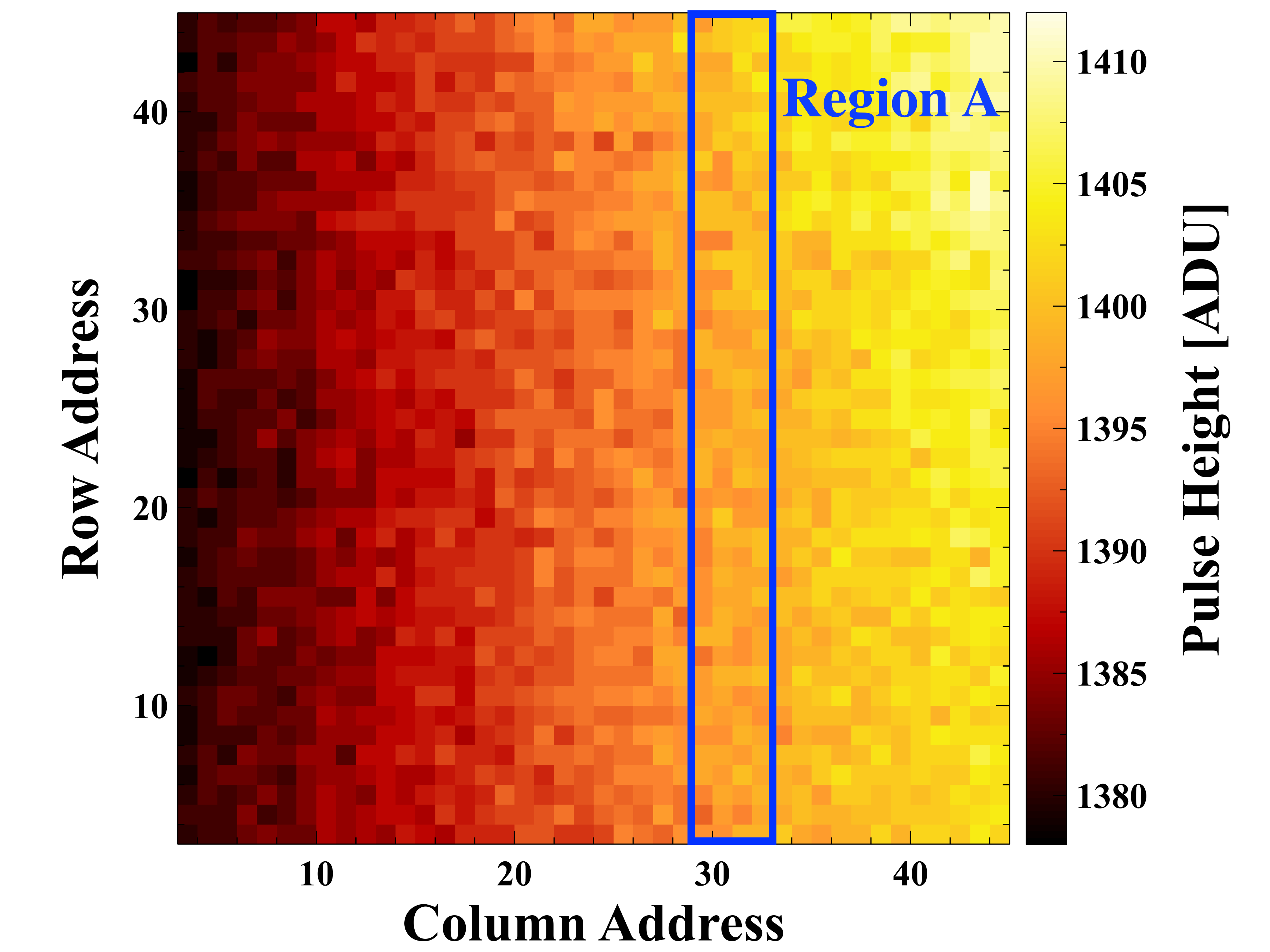}
\caption{Pedestal map obtained in the Frame readout mode. 
The blue rectangle region ``Region A'' is extracted for the evaluation of the trigger performance.}
\label{fig:Pedestal_Map}
\end{figure}

The cause of the non-uniformity in the counts maps is the non-uniformity of the pedestal voltages at the input to the comparator circuit as described below. 
Assuming that an offset voltage exists at the input to the comparator circuit, it will be triggered by a signal voltage with the offset added to it. 
In order to examine the offset voltage and its uniformity, we obtained a pedestal map as presented in Fig.~\ref{fig:Pedestal_Map}, where we show means of raw pulse height values of each pixel.
The measurement was performed in the Frame readout mode without X-ray irradiation.
The map indicates the pedestal level, which would correspond to the offset of the comparator input and of the CDS output, tends to be lower in pixels on the left-hand side. 
When the threshold is set to a voltage close to incoming X-ray energy, we can expect a situation where pixels with larger offsets fire triggers but those with smaller offsets do not.
This would be the explanation for the gradation we observed in the counts maps (Fig.~\ref{fig:Trigger_Map}).
The peak-to-peak variation in the pedestal non-uniformity (Fig. \ref{fig:Pedestal_Map}) is $\sim 35 ~\mathrm{ADU}$, which corresponds to X-ray energy of $\sim 800 ~\mathrm{eV}$ calculated by the linear relationship between ADU and X-ray energy ($45.0~\mathrm{ADU/keV}$) obtained from X-ray spectra.
This value is translated to $\sim 10 ~\mathrm{mV}$ at the output from the CSA circuit by using $1 ~\mathrm{ADU} = 488~\mathrm{\mu V}$, the mean ionization energy per electron-hole pair in Si ($3.65 ~\mathrm{eV}$), and the second-stage amplifier with a gain of $1.8$.

The cause of the non-uniformity of the pedestal voltages is unknown and still under investigation at this moment. 
The comparator circuit itself does not cause the non-uniformity because the non-uniformity of the pedestal voltages is observed even in the Frame readout mode, in which the circuit is not used. 
The dark current, which is evaluated based on exposure-time dependence of the pedestal level,
is not the cause since the pixel-to-pixel variation of the dark current is $\sim 5 ~ e^{-}/\mathrm{msec/pixel}$, which is equivalent to X-ray energy of $\sim 20 ~\mathrm{eV}$.
Here the exposure time in the Frame readout mode is $1 ~\mathrm{msec}$, and the maximum exposure time in the Event-Driven readout mode is also $1 ~\mathrm{msec}$.
An offset voltage in the output from the CSA circuit is considered not to cause the non-uniformity shown in Fig.~\ref{fig:Pedestal_Map}, because the offset should be canceled by the CDS circuit located at the input of the comparator circuit.
In conclusion, the cause of the non-uniformity is still unknown. 
Although we think that the pixel circuit is likely to be the cause of the problem, we will examine the overall design of the XRPIX6E device, such as peripheral circuits, wirings, and the buried wells in the sensor layer to explore the cause.


\subsection{Characterization in a limited region}
\label{subsec:Characterization in a Limited Region}
\begin{figure}[tbp]
\centering
\includegraphics[width=.95\columnwidth]{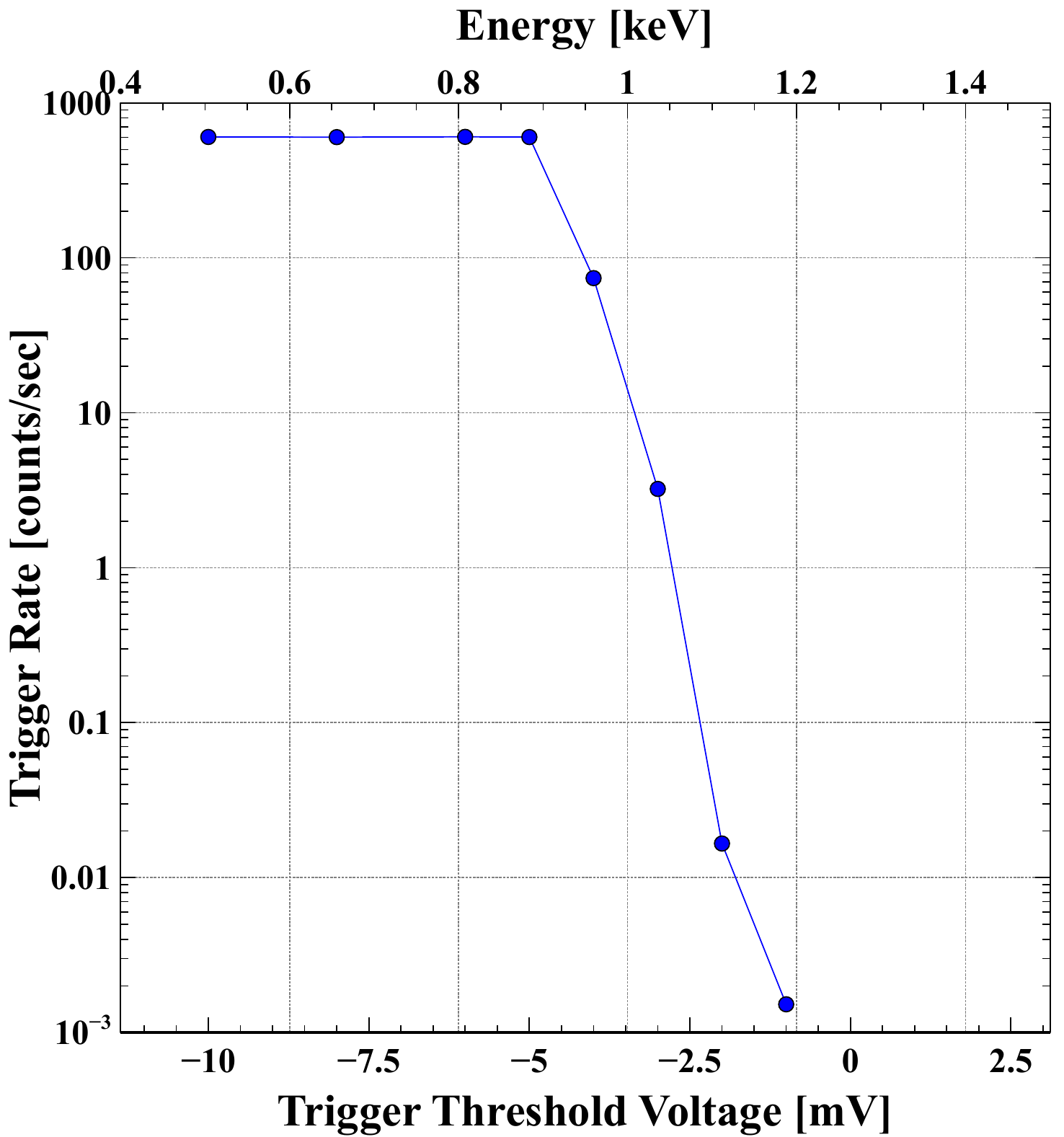}
\caption{Noise trigger rate as a function of trigger threshold voltage ($V_{\mathrm{tr}}$).}
\label{fig:NoiseRate_vs_Vtr}
\end{figure}

In the following, we evaluate the performance of the Event-Driven readout mode by limiting the data acquisition region to ``Region A'' shown in  Fig.~\ref{fig:Pedestal_Map} to minimize the effect of the region-to-region pedestal variation.
The variation in Region A is equivalent to noise charge of $\sim 11~e^{-}$ in rms.
Fig.~\ref{fig:NoiseRate_vs_Vtr} shows the trigger rates as a function of the trigger threshold voltage without X-ray illumination,
 where noise fires triggers with a lower threshold voltage. 
The lowest threshold voltage that can avoid such false triggers is $V_{\mathrm{tr}} = -2 ~\mathrm{mV}$, which corresponds to the lowest detectable X-ray energy in the Event-Driven readout mode. 
Adopting the threshold voltage of $V_{\mathrm{tr}} = -2 ~\mathrm{mV}$, we observed the characteristic X-rays with the same experimental setup as used for the evaluation of the QEs in the Frame readout mode. 
Note that no data were taken for Ti-K X-rays due to the limitation of experimental time. 
We successfully obtained Al-K spectrum in the Event-Driven readout mode for the first time in the XRPIX series (Fig.~\ref{fig:spectra_Al_F}~(b)), while F-K X-rays were not detected. 

\begin{figure}[tbp]
\centering
\includegraphics[width=.95\columnwidth]{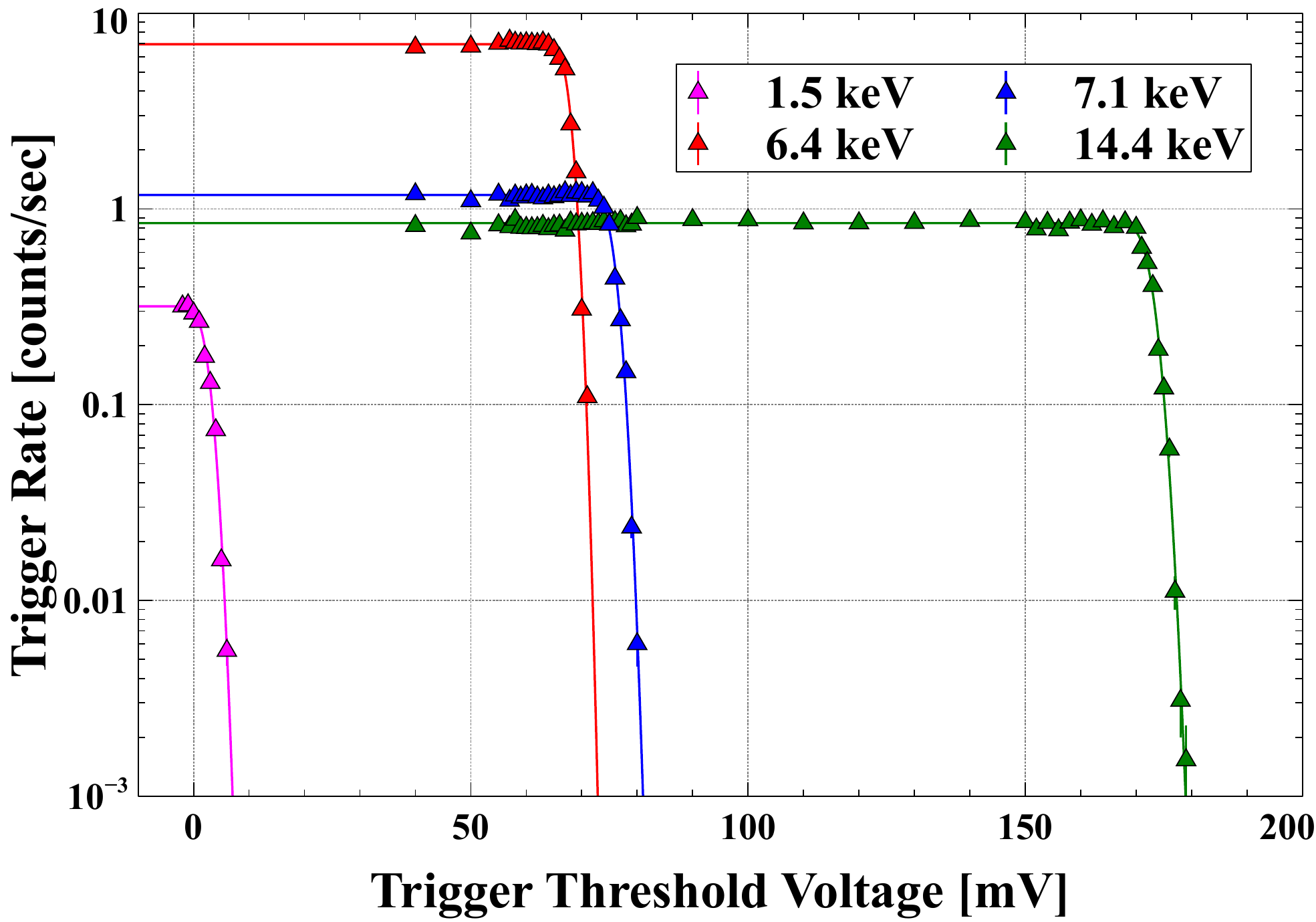}
\caption{Trigger output rates of monochromatic lines as a function of trigger threshold voltage ($V_{\mathrm{tr}}$). 
	The red, blue, green plots represent $6.4, 7.1, 14.4 ~ \mathrm{keV}$, respectively obtained with a $^{57} \mathrm{Co}$ radioisotope. 
	The magenta plots show the data of $1.5 ~ \mathrm{keV}$ X-rays corresponding to Al-K line. 
	Each data set is fitted with a complementary error function shown with the curves in its color.}
\label{fig:TriggerRate_vs_Vtr}
\end{figure}

\begin{figure}[htbp]
\centering
\includegraphics[width=.95\columnwidth]{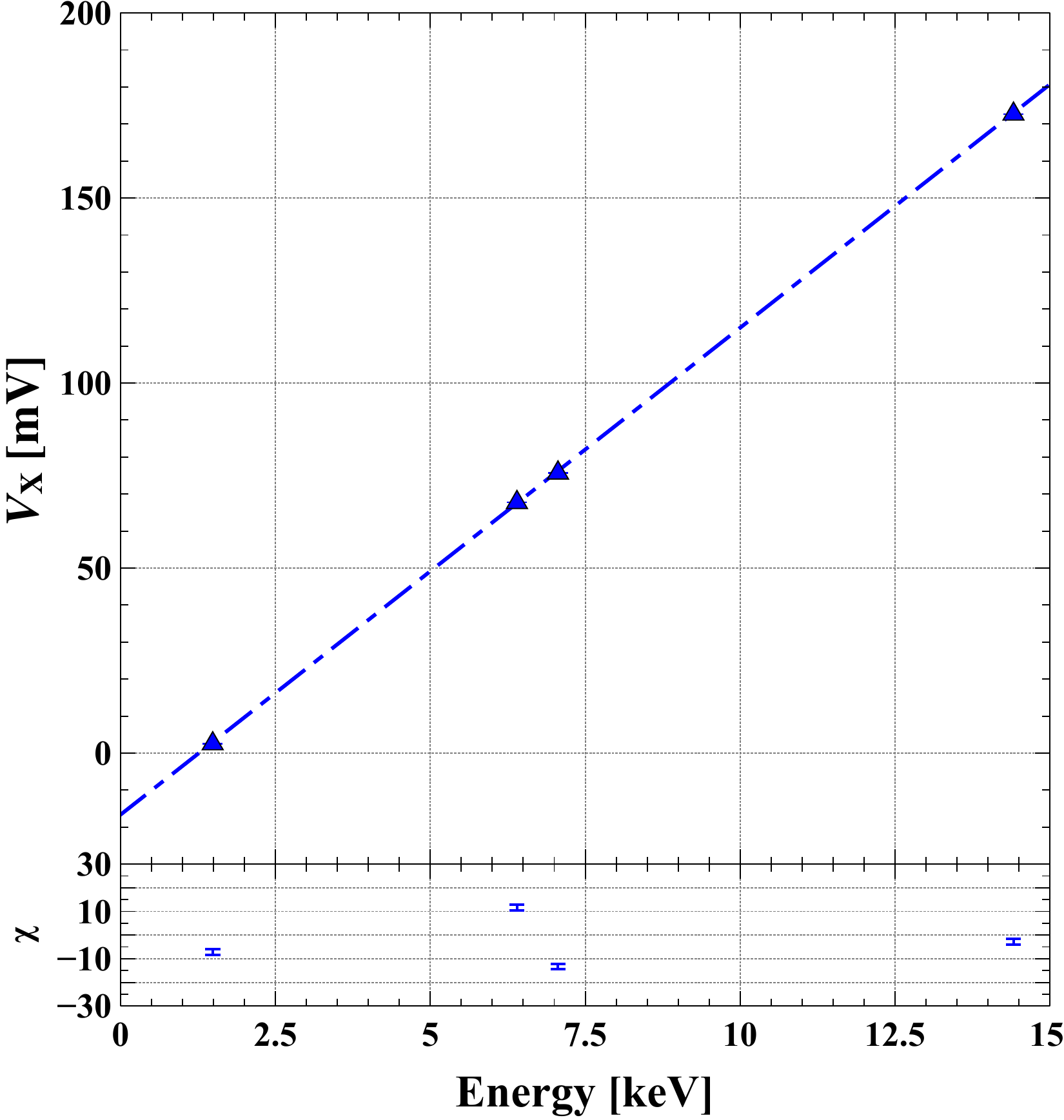}
\caption{Liner relationship between $V_{\mathrm{X}}$ and energy.}
\label{fig:Vtr_vs_E}
\end{figure}

Fig.~\ref{fig:TriggerRate_vs_Vtr} shows the trigger rates as a function of threshold voltage ($V_{\mathrm{tr}}$) for each characteristic X-ray line. 
We show $V_{\mathrm{X}}$ as a function of X-ray energy in Fig.~\ref{fig:Vtr_vs_E} obtained with the so-called ``S-curve method'' (e.g.\cite{Kraft et al.(2009)}). 
For each characteristic X-ray line, the data points were fitted with a complementary error function, or the integral of Gaussian defined as 
\[
\label{eq:erf}
\mathrm{Trigger\ Rate} \propto \int^\infty_{V_\mathrm{tr}} \exp\left(-\frac{\left(V-V_\mathrm{X}\right)^2}{2w_\mathrm{X}^2}\right)dV. 
\]
Here the inflection point $V_\mathrm{X}$ is the voltage corresponding to the energy of the characteristic X-ray. 
The parameter $w_\mathrm{X}$ is related to the noises of the comparator and the CSA circuits, the Fano noise, effects from the dark current, and the pixel-to-pixel variations of the gain and pedestal.
We hereafter refer to $w_\mathrm{X}$ as ``threshold energy resolution.'' 
A good linear correlation between $V_{\mathrm{X}}$ and X-ray energy is seen and its gain of $48.0~\mathrm{\mu V}/e^{-}$ agrees well with the node sensitivity of $44.6~\mathrm{\mu V}/e^{-}$.
According to the correlation, the lowest available threshold voltage of $V_{\mathrm{tr}} = -2 ~\mathrm{mV}$ is translated to $1.1~ \mathrm{keV}$, which is consistent with the experimental result that F-K X-rays at $0.68~ \mathrm{keV}$ were not detected.
The obtained value, $1.1~\mathrm{keV}$ is still higher than the requirement of the FORCE mission of $1.0~\mathrm{keV}$, indicating that further improvements are necessary. 

The QEs for Al-K, Fe-K and the 14.4-$\mathrm{keV}$ line in the Event-Driven readout mode are compared with those in the Frame readout mode in Fig.~\ref{fig:QE}. 
There is no difference between the two modes for Fe-K X-rays and the 14.4-$\mathrm{keV}$ line. 
On the other hand, the QE of Al-K in the Event-Driven readout mode is $\sim 20\%$ lower than that in the Frame readout mode. 
The trigger threshold energy in the Event-Driven readout mode corresponds to the event threshold of the Frame readout mode, and its value ($1.1~\mathrm{keV}$) is significantly higher than the event threshold ($0.45~\mathrm{keV}$) in the Frame readout mode. 
This would be the reason for the lower QE for Al-K X-rays in the Event-Driven readout mode. 
Since the trigger threshold energy is higher than a half of the Al-K X-ray energy of $1.5~\mathrm{keV}$, some of Al-K X-ray events whose signal charge is split into two or more pixels are not detected. 
Hence, it is necessary to lower the trigger threshold energy in order to obtain the same QEs for low-energy X-rays as those in the Frame readout mode. 

\subsection{Noise performance of trigger circuit}
\begin{figure}[tbp]
\centering
\includegraphics[width=.95\columnwidth]{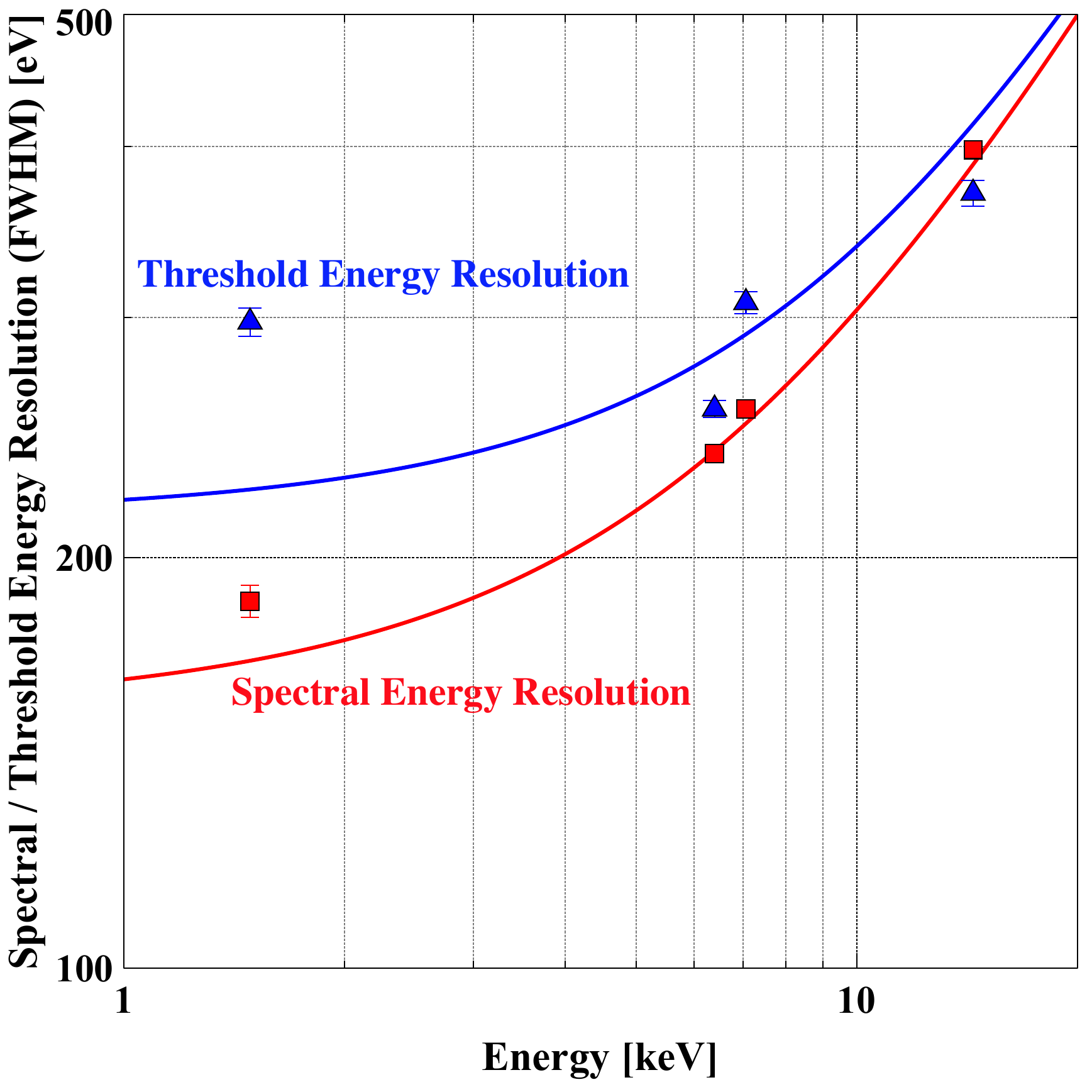}
\caption{Spectral and threshold energy resolutions in the Event-Driven readout mode. 
}
\label{fig:FWHM_Comparison}
\end{figure}

Fig.~\ref{fig:FWHM_Comparison} shows the threshold energy resolution of the trigger output function and the spectral energy resolution obtained with the X-ray spectra of the single-pixel events from Region A in the Event-Driven readout mode. 
They are presumed to be described by the following function
\[
\label{eq:FWHM}
\mathrm{\Delta E} = 2.355\ W \sqrt{\frac{FE}{W} + {\sigma_\mathrm{inde}}^{2} + \left(\frac{E\sigma_\mathrm{gain}}{W}\right)^2},
\]
where $\Delta E$, $E$, $F$, $W$, $\sigma_\mathrm{inde}$ and $\sigma_\mathrm{gain}$ are the threshold or spectral energy resolution in FWHM, the incident X-ray energy, the Fano factor (0.12), the mean ionization energy per electron-hole pair in Si ($3.65 ~\mathrm{eV}$),
the energy independent noise in rms, and the pixel-to-pixel relative gain variation in rms, respectively. 
While the noise at the output from the CSA circuit dominates the energy independent noise for the spectral energy resolution, the noise of the comparator circuit, the non-uniformity of the pedestal, and the dark current contribute as well for the threshold energy resolution. 
As shown in Fig.~\ref{fig:FWHM_Comparison}, we estimated the energy independent noise, i.e. the noise of the CSA circuit operating in the Event-Driven readout mode, and the pixel-to-pixel relative gain variation for the spectral energy resolution to be $\sim 18~e^{-}$ (rms) and $\sim 0.9\%$ (rms), respectively. 
We also estimated the energy independent noise to be $\sim 25~e^{-}$ (rms) for the threshold energy resolution, in which we adopted the same pixel-to-pixel relative gain variation as that obtained for the spectral energy resolution. 

We presume that the lowest available threshold energy ($1.1~ \mathrm{keV}$) for triggers in the Event-Driven readout mode depends on the energy independent noise and the equivalent noise due to the non-uniformity of the pedestal ($\sim 25~e^{-}$ in rms). 
Assuming that we can successfully reduce the non-uniformity sufficiently in future, the remaining energy independent noise will be $\sim 22~e^{-}$ (rms), which can be translated to the lowest detectable X-ray energy of $1.0~\mathrm{keV}$. 
This remaining energy independent noise consists of the CSA circuit noise and the one related to the trigger output function including the comparator circuit. 
The noise related to the trigger output function is estimated to be $\sim 13~e^{-}$ (rms) by subtracting the CSA circuit noise of $\sim 18~e^{-}$ (rms) from the $\sim 22~e^{-}$ (rms) noise. 
Following the discussion on the event threshold in the Frame readout mode given in section~\ref{sec:QE}, it is still necessary to lower the energy threshold even further to detect multiple-pixel events at $1.0~\mathrm{keV}$. 
For example, the energy independent noise of $\sim 11~e^{-}$ (rms) is necessary to detect two-pixel events of $1.0~ \mathrm{keV}$ X-rays efficiently. 
It requires further reduction of both the noise in the CSA circuit and the noise related to the trigger output function.

\section{Conclusion}
\label{sec:conclusion}
We have presented the low-energy X-ray performance of the XRPIX6E device with the PDD structure.
The PDD structure realizes the low readout noise by suppressing the electrical interference between the sensor layer and the circuit layer.
F-K X-rays ($0.68~\mathrm{keV}$) and Al-K X-rays ($1.5~\mathrm{keV}$) were clearly detected with backside illumination measured in the Frame readout mode, and the effective thickness of the dead layer on the backside is estimated to be $0.60 \pm 0.15 ~\mathrm{\mu m}$ from the quantum efficiencies obtained with the low-energy X-rays.
Though Al-K X-rays were successfully detected in the Event-Driven readout mode, F-K X-rays were not.
The lowest detectable X-ray energy was $1.1~ \mathrm{keV}$ for a limited region where the non-uniformity of the pedestal voltages is almost negligible.
This high energy threshold in the Event-Driven readout mode is considered to be caused by 
the CSA circuit noise of $\sim 18~e^{-}$ (rms) and the noise related to the trigger output function of $\sim 13~e^{-}$ (rms).
Reducing the sum of these noises to $\sim 11~e^{-}$ (rms) is necessary to satisfy the requirement of the FORCE mission to detect 1-keV X-rays with high quantum efficiency in the Event-Driven readout mode.

\section*{Acknowledgement}
We acknowledge the valuable advice and the manufactures of XRPIXs by the personnel of LAPIS Semiconductor Co., Ltd.
This study was supported by the Japan Society for the Promotion of Science (JSPS) KAKENHI, Japan Grant-in-Aid for Scientific Research on Innovative Areas 25109004 (T.G.T. \& T.T.), 25109003 (S.K.) and 25109002 (Y.A.), 
Grant-in-Aid for Scientific Research (A) 15H02090 (T.G.T.), 
Grant-in-Aid for Challenging Exploratory Research 26610047 (T.G.T.), 
Grant-in-Aid for Young Scientists (B) 15K17648 (A.T.) and Grant-in-Aid for JSPS Fellows 15J01842 (H.M.).
This study was also supported by the VLSI Design and Education Center (VDEC), the University of Tokyo in collaboration with Cadence Design Systems, Inc., and Mentor Graphics, Inc.





\begin{thebibliography}{99}


\bibitem{Mori et al.(2016)}
K. Mori, T. G. Tsuru, K. Nakazawa et al., 
	A broadband x-ray imaging spectroscopy with high-angular resolution: the FORCE mission,
	in Proc. SPIE, 9905 (2016) 99051O.

\bibitem{Nakazawa et al.(2018) FORCE}
K. Nakazawa, K. Mori, T. G. Tsuru et al., 
	The FORCE mission: science aim and instrument parameter 
	for broadband x-ray imaging spectroscopy with good angular resolution, 
	in Proc. SPIE, 10699 (2018) 106992D.

\bibitem{Zhang et al.(2019)}
W. W. Zhang, K. D. Allgood, M. P. Biskach et al., 
	Next generation x-ray optics for astronomy: high resolution, lightweight, and low cost, 
	in Proc. SPIE, 11119 (2019) 1111907.

\bibitem{Nakazawa et al.(2018) HXI}
K. Nakazawa, G. Sato, M. Kokubun, et al.,
	Hard x-ray imager onboard Hitomi (ASTRO-H),
	Journal of Astronomical Telescopes, Instruments, and Systems, 4 (2018) 021409.

\bibitem{Tsuru et al.(2018)}
T. G. Tsuru, H. Hayashi, K. Tachibana et al., 
	Kyoto's event-driven x-ray astronomy SOI pixel sensor for the FORCE mission, 
	in Proc. SPIE, 10709 (2018) 107090H.

\bibitem{Arai et al.(2011)}
Y. Arai, T. Miyoshi, Y. Unno et al., 
	Development of SOI pixel process technology, 
	Nuclear Instruments and Methods in Physics Research Section A, 636 (2011) S31.





\bibitem{Itou et al.(2016)}
M. Itou, T. G. Tsuru, T. Tanaka et al., 
	X-ray Performance of Back-Side Illuminated Type of Kyoto's X-ray Astronomical SOI Pixel Sensor, XRPIX, 
	Nuclear Instruments and Methods in Physics Research Section A, 831 (2016) 55.

\bibitem{Negishi et al.(2019)}
K. Negishi, T. Kohmura, K. Hagino et al., 
	X-ray response evaluation in subpixel level for X-ray SOI pixel detectors, 
	Nuclear Instruments and Methods in Physics Research Section A, 924 (2019) 462.

\bibitem{Kamehama et al.(2018)}
H. Kamehama, S. Kawahito, S. Shrestha et al., A Low-Noise X-ray Astronomical Silicon-On-Insulator Pixel Detector Using a Pinned Depleted Diode Structure, Sensors, 18 (2018) 27.

\bibitem{Harada et al.(2019)}
S. Harada, T. G. Tsuru, T. Tanaka et al. ,Performance of the Silicon-On-Insulator pixel sensor for X-ray astronomy, XRPIX6E, equipped with pinned depleted diode structure, Nuclear Instruments and Methods in Physics Research Section A, 924 (2019) 468.

\bibitem{Kayama et al.(2019)}
K. Kayama, T. G. Tsuru, T. Tanaka et al. ,Subpixel response of SOI pixel sensor for X-ray astronomy with pinned depleted diode: first result from mesh experiment, Journal of Instrumentation, 14 (2019) C06005.


\bibitem{Takeda et al.(2013)}
A. Takeda, Y. Arai, S. G. Ryu et al., Design and Evaluation of an SOI Pixel Sensor for Trigger-Driven X-ray Readout, IEEE Transaction on Nuclear Science, 60 (2013) 586.

\bibitem{Ryu et al.(2011)}
S. G. Ryu, T. G. Tsuru, S. Nakashima et al., First Performance Evaluation of an X-Ray SOI Pixel Sensor for Imaging Spectroscopy and Intra-Pixel Trigger, IEEE Transactions on Nuclear Science, 58 (2011) 2528.

\bibitem{Kraft et al.(2009)}
P. Kraft, A. Bergamaschi, C. Bronnimann, et al., Characterization and Calibration of PILATUS Detectors, IEEE Transactions on Nuclear Science, 56 (2009) 758.











\end{thebibliography}


\end{document}